\documentclass[reprint,superscriptaddress,amsmath,amssymb,aps,prl]{revtex4-1}

\usepackage{graphicx}
\usepackage{dcolumn}
\usepackage{bm}
\begin{document}

\preprint{APS/123-QED}

\title{Field Induced Positional Shift of Bloch Electrons and Its Dynamical Implications}

\author{Yang Gao}
\affiliation{Department of Physics, The University of Texas at Austin, Austin, Texas 78712, USA}
 
\author{Shengyuan A. Yang}
\affiliation{Engineering Product Development, Singapore University of Technology and Design, Singapore 138682, Singapore}
 
\author{Qian Niu}
\affiliation{Department of Physics, The University of Texas at Austin, Austin, Texas 78712, USA}
\affiliation{International Center for Quantum Materials, Peking University, Beijing 100871, China}

\date{\today}

\begin{abstract}
We derive the field correction to the Berry curvature of Bloch electrons, which can be traced back to a positional shift due to the interband mixing induced by external electromagnetic fields. The resulting semiclassical dynamics is accurate to second order in the fields, in the same form as before, provided that the wave packet energy is derived up to the same order. As applications, we discuss the orbital magnetoelectric polarizability and predict nonlinear anomalous Hall effects.
\begin{description}
\item[PACS numbers]72.15.-v, 72.10.Bg, 73.43.-f, 85.75.-d

\end{description}
\end{abstract}


\maketitle

The response of Bloch electrons to external fields has been a central topic in solid state physics. Due to the Berry curvature of Bloch states, the semiclassical dynamics acquires a non-canonical structure \cite{Chang2008, Xiao2005, littlejohn1991}. This is manifested as an anomalous velocity and a modification of the phase-space density of states, with important consequences on the thermodynamic and transport properties \cite{Xiao2005, Xiao2010, Sun1999, Chang1995}. Together with a first order correction to the band energy due to the orbital magnetic moment, the Berry curvature provides the essential ingredient for a full theory of the electron response to first order in external fields.

However, response functions such as electric polarizability, magnetic susceptibility, and magnetoelectric polarizability would require a theory that is accurate up to second order in external fields. The difficulty in establishing this type of theory originates from the unboundedness of the perturbative Hamiltonian. Blount
 pioneered the work of systematically extending semiclassical theory up to second order by using phase space quantum mechanics \cite{Blount1962a}. However, his method uses variables which are not fully gauge invariant with respect to the phase choice in the basis Bloch states, rendering it difficult to understand the physical meaning of his results, especially so because of some unresolved gauge issues.
\cite{foot1}.

In this letter, we present a second order semiclassical theory for Bloch electrons under uniform electromagnetic fields in terms of physical position and crystal momentum which are fully gauge invariant. A central concept is a gauge-invariant positional shift due to field induced interband mixing. It leads to a field correction to the Berry curvature, and modifies the relationship between the physical position and crystal momentum with the canonical ones. However, to our surprise and delight, the resulting equations of motion up to second order still retain the same form as in the first order theory, provided that the band energy is also corrected to second order in the fields.

The field induced positional shift of Bloch electrons has profound implications. It is solely responsible for the cross-gap part of the orbital magnetoelectric polarizability \cite{FKM, QHZ,Moore2007,Roy2009,Vander2009, Vander2010}. Moreover, its resulting field correction to the Berry curvature also leads to nonlinear anomalous Hall effects, with a Hall conductivity proportional to external electric or magnetic field. The electric nonlinear anomalous Hall conductivity is intimately related to the orbital magnetoelectric polarizability, and requires the system to have both time reversal and spatial inversion symmetry breaking. The magneto nonlinear anomalous Hall effect does not have such symmetry restrictions, and it competes with the ordinary Hall effect in relatively dirty samples. Besides these two well-analysed applications, our complete second order semiclassical theory also provides straightforward methods to evaluate
magnetic susceptibility, electric polarizability, magnetoresistance, intrinsic thermoelectric current, etc. All results from our theory can be easily evaluated in the first principle calculations to be compared with experiments.

{\it Positional Shift}.---The basic idea of the semiclassical theory is to study the evolution of a wave packet constructed from a single Bloch band (labelled by subscript 0 and we focus on the Abelian case for simplicity). One starts from the local Hamiltonian obtained from the full quantum Hamiltonian by evaluating the gauge potentials at the center of mass position $\bm r_c$ of the wave packet, $
\hat{H}_c(\hat{\bm p},\hat{\bm q})=\hat{H}_0(\hat{\bm p}+{1\over 2}\bm B\times \bm r_c,\hat{\bm q})+\bm E\cdot \bm r_c\,,$
where $\hat{H}_0(\hat{\bm p},\hat{\bm q})$ is the unperturbed Hamiltonian, $\hat{\bm p}$ and $\hat{\bm q}$ are momentum and position operators, and we have set $e=\hbar=1$ to simplify notations. In the first order semiclassical theory, the wave packet is constructed by superposing the Bloch eigenstates $e^{i\bm q\cdot \bm p}|u_0 (\bm p+{1\over 2}\bm B\times \bm r_c)\rangle$ of this local Hamiltonian with crystal momentum $\bm p$ centered around some point $\bm p_c$, satisfying the self-consistency requirement that the position expectation value of the wave packet coincides with the presumed value $\bm r_c$. A non-canonical geometry of the semiclassical dynamics emerges in the equations of motion \cite{Sun1999} for the physical position $\bm r_c$ and the gauge invariant crystal momentum $\bm k_c=\bm p_c+{1\over 2}\bm B\times \bm r_c$. This involves the Berry curvature and the orbital magnetic moment in the unperturbed Bloch band, which are necessary and sufficient to make the dynamics and its quantum extension accurate to first order in external fields \cite{Chang2008}.

In the second order theory, we need the first order correction $|u_0^\prime\rangle$ to the unperturbed band $|u_0\rangle$ by the gradient perturbation $\hat{H}^\prime={1\over 4m}\bm B\cdot [(\hat{\bm q}-\bm r_c)\times \hat{\bm V}-\hat{\bm V}\times (\hat{\bm q}-\bm r_c)]+\bm E\cdot (\hat{\bm q}-\bm r_c)$ to the local Hamiltonian $\hat{H}_c$, where $\hat{\bm V}=-i[\hat{\bm q},\hat{H}_0]$ is the velocity operator. Based on the perturbed band, the wave packet now acquires a shift in its center of mass position given by $\bm a^\prime=\langle u_0|i\partial_{\bm p} |u_0^\prime\rangle+c.c.$. It corresponds to a first order correction to the Berry connection $\bm a=\langle u_0|i\partial_{\bm p}|u_0\rangle$ of the unperturbed band, but is gauge invariant as can be easily checked by using the orthogonality between $|u_0^\prime\rangle$ and $|u_0\rangle$. Therefore it is a physical quantity and represents the shift of the wave packet center due to external fields. Indeed, $\bm a'$ transforms as a spatial vector under symmetry operations, e.g. it is odd under spatial inversion and even under time reversal. Furthermore, it should be noted that $\bm a^\prime$ is a periodic function of the lattice, which does not cause any macroscopic charge density gradient, hence it will not affect the electron's chemical potential profile. This positional shift is the central concept of our theory.

Following the standard perturbation scheme \cite{suppl}, we can derive the following explicit formula for the positional shift (of band 0) in terms of the unperturbed Bloch bands and external fields:
\begin{equation}\label{newa}
a_i^\prime=F_{ij}B_j +G_{ij}E_j\,,
\end{equation}
where
\begin{equation}\label{FG}
F_{ij}=\mathfrak{Im}\sum_{n\neq 0} {(V_i)_{0n} (\omega_j)_{n0}\over (\varepsilon_0-\varepsilon_n)^2},\;G_{ij}=2\mathfrak{Re}\sum_{n\neq 0}{(V_i)_{0n} (V_j)_{n0} \over (\varepsilon_0-\varepsilon_n)^3}\,,
\end{equation}
with ${\bm \omega}_{n0}$ defined as
\begin{equation}
(\omega_j)_{n0}=-i\epsilon_{jk\ell}\sum_{m\neq 0}  {[(V_k)_{nm}+(v_k)\delta_{nm}](V_\ell)_{m0}\over \varepsilon_m-\varepsilon_0}\,.
\end{equation}
Here, $\varepsilon_0$, $\varepsilon_n$ and $\varepsilon_m$ are band dispersions for band $0,n$, and $m$, respectively, $i,\,j,\,k,\,\ell$ refer to the spatial components, $\epsilon_{jk\ell}$ is the anti-symmetric tensor, $v_\ell=\partial_{p_\ell} \varepsilon_0$ is the group velocity of band 0,  and $(V_k)_{nm}$ is the matrix element of the velocity operator. Here and hereafter, summation is implied over repeated spatial indices. Since $\bm a^\prime$ contains the interband velocity, it is not a single band property. All the quantities in Eq.(\ref{FG}) can be readily evaluated in first principle calculations.

To illustrate the positional shift, we consider a generic two-band model Hamiltonian with 
\begin{equation}\label{2band}
\hat{H}_0=h_0+\bm h\cdot \bm \sigma\,
\end{equation} 
where $\bm \sigma$ is the vector of Pauli matrices and $h$\rq{}s have arbitrary dependence on the crystal momentum.
The energy band dispersion is $\varepsilon_{\pm}=h_0\pm h$. Assume the two bands are fully gapped with $h\neq 0$. The positional shift for the lower band can be calculated from Eq.(\ref{newa}) and (\ref{FG}) with
\begin{equation}\label{dipole}
F_{ij}=-{g_{ik} \epsilon_{k\ell j} \partial_{p_\ell} h_0\over 4h}-{1\over 8} \epsilon_{jk \ell}\mathit \Gamma_{\ell k i},\quad G_{ij}=-{1\over 4h}g_{ij},
\end{equation}
where $g_{ik}=\partial_{p_i} \bm n \cdot \partial_{p_k} \bm n$ (with $\bm n=\bm h/h$) is the quantum metric of the band,  and $\mathit \Gamma_{\ell k i}={1\over 2}(\partial_{p_i}g_{\ell k}+\partial_{p_k}g_{\ell i}-\partial_{p_\ell}g_{ki})$ is the corresponding Christoffel symbol \cite{anandan1990}. Like the Berry curvature, the quantum metric is also a geometric physical quantity, which defines the infinitesimal distance in the Hilbert space on the Brillouin zone. Meanwhile the Christoffel symbol defines the affine geometry of the Brillouin zone \cite{anandan1990}. They together make the Brillouin zone a Riemannian manifold. It has been proposed that the quantum metric could be probed by measuring the current noise spectrum \cite{Neupert2013}. Our result shows that $g$ and $\mathit\Gamma$ are also closely connected with the positional shift, hence might be probed in second order effects.

{\it Second Order Semiclassical Theory.}---To see how the positional shift enters the second order semiclassical dynamics, we derive the effective Lagrangian for the wave packet dynamics \cite{suppl}:
\begin{align}\label{lag}
L
&=-(\bm r_c-\bm a-\bm a^\prime)\cdot \dot{\bm k}_c-{1\over 2}\bm B\times \bm r_c\cdot \dot{\bm r}_c-\tilde{\varepsilon}\,,
\end{align}
where $\bm k_c=\bm p_c+{1\over 2}\bm B\times \bm r_c$ is the gauge invariant crystal momentum and $\tilde{\varepsilon}$ is the semiclassical energy accurate to second order (see discussion of $\tilde{\varepsilon}$ in supplemental materials \cite{suppl}). In deriving $L$, the crystal momentum $\bm p$ for each Bloch component of the wave packet has been integrated out, so quantities such as $\bm a$, $\bm a^\prime$ here are now functions of $\bm k_c$, instead of $(\bm p+{1\over 2}\bm B\times \bm r_c)$. 

One direct consequence of the positional shift is that the Berry curvature $\bm{\mathit{\Omega}}$ now acquires a field correction given by $\bm{\mathit{\Omega}}^\prime=\bm \partial\times \bm a^\prime$. (Here and hereafter, the partial derivative $\bm \partial$ is with respect to $\bm k_c$ unless being explicitly pointed out otherwise.) Surprisingly, with this modified Berry curvature $\tilde{\bm{\mathit{\Omega}}}=\bm{\mathit{\Omega}}+\bm{ \mathit{\Omega}}^\prime$ and the second order wave packet energy $\tilde{\varepsilon}$, the Euler-Lagrange equations of motion have the same form as in the first order theory \cite{Sun1999,suppl}:
\begin{align}
\label{rdot} \dot{\bm r}_c&={\partial \tilde{\varepsilon}\over \partial \bm k_c}-\dot{\bm k}_c\times \bm \tilde{\bm{\mathit{\Omega}}},\\
\label{kdot} \dot{\bm k}_c&=-\bm E-\dot{\bm r}_c \times \bm B\,.
\end{align}

The force equation remains the same as before, and the velocity equation now involves the modified quantities with field corrections. Similar to the first order theory \cite{Xiao2005}, the phase space density of states has a correction factor $\mathcal{D}=1+\bm B\cdot \tilde{\bm{\mathit{\Omega}}}$, which is now accurate to second order in the fields. It is important to note that even though the Berry curvature is corrected, the Chern number, which is the integral of Berry curvature over the Brillouin zone, is not affected. This is because $\bm a^\prime$ is well defined and periodic in the Brillouin zone hence the integral of its curl over the entire zone necessarily vanishes.

The positional shift also modifies the relationship between the physical variables ($\bm r_c$, $\bm k_c$) and the canonical variables $(\bm q, \bm p)$ \cite{Chang2008}. These relations are important for re-quantizing the semiclassical theory, and are directly related to physical quantities such as charge polarization discussed in the following. As detailed in the supplemental materials \cite{suppl}, we follow the procedure in Ref.\cite{Chang2008} and obtain that 
\begin{align}\label{rc}
\bm r_c&=\bm q+\bm a+{1\over 2}(\bm B\times \bm a\cdot \bm \partial_{\bm p})\bm a+{1\over 2}\bm{\mathit{\Omega}}\times (\bm B\times \bm a)+\bm a^\prime\,,\\
\label{kc}\bm k_c&=\bm p+{1\over 2}\bm B\times \bm q+\bm B\times (\bm r_c-\bm q)\,.
\end{align}
Note that in Eq.(\ref{rc}) and (\ref{kc}), the argument for $\bm a$ and $\bm a^\prime$ is $\bm p+{1\over 2}\bm B\times \bm q$.
It was previously thought that the first four terms on the right hand side of Eq.(\ref{rc}) would be sufficient to first order in the fields \cite{Chang2008}, but now one observes that this is incomplete without the positional shift $\bm a'$.

{\it Orbital magnetoelectric polarizability.}---In the absence of external fields, the polarization from electrons in a filled band is given by the integral of the Berry connection $\bm a$ \cite{Smith1993, Resta1994}. From a semiclassical point of view, a magnetic field $\bm B$ modifies this formula in two ways: (1) the density of states in the integral should contain the factor $\mathcal{D}=1+\bm B\cdot \tilde{\bm{\mathit{\Omega}}}$; (2) the Berry connection $\bm a$ should be replaced by $\bm a+(\bm B\times \bm a\cdot \bm \partial_{\bm p})\bm a/2+\bm{\mathit{\Omega}}\times (\bm B\times \bm a)/2+\bm a^\prime$ according to the relationship between the physical position and the canonical position in Eq.(\ref{rc}). Combining these modifications up to first order in the magnetic field and rewriting them in terms of the gauge invariant crystal momentum $\bm k_c$, we obtain the polarization $\bm P^\prime$ that is first order in $\bm B$ field:
\begin{equation}\label{me}
\bm P^\prime=-\int{d^3k\over (2\pi)^3}\left[{1\over 2}(\bm{\mathit{\Omega}}\cdot\bm a)\,\bm B+\bm a^\prime\right]\,,
\end{equation}
where the integral is over the Brillouin zone and we drop the subscript $c$ of momentum $k$ for simple notations.

The first term in Eq.(\ref{me}) is the Abelian Chern-Simons form, which plays a central role in the classification of three dimensional topological insulators \cite{QHZ, Zhang2011, Kane2010, Vish2010, Bern2011, Fu2007}. It only involves the Berry connection and Berry curvature of the unperturbed band, and can be derived within the framework of the first order semiclassical theory \cite{Vander2009, Xiao2009}. The additional term from the field-induced band mixing was envisioned in Ref.\cite{Xiao2009}, but its vadidation had to wait for a full quantum perturbation treatment in Ref.\cite{Vander2010}. We now see that this additional term actually comes in the nice form of the positional shift integrated over the Brillouin zone.
Our result agrees exactly with the full quantum result, confirming the reliability of our semiclassical theory.

\begin{figure}[!htp]
\setlength{\abovecaptionskip}{1pt}
\setlength{\belowcaptionskip}{1pt}
\includegraphics[width=1.4in]{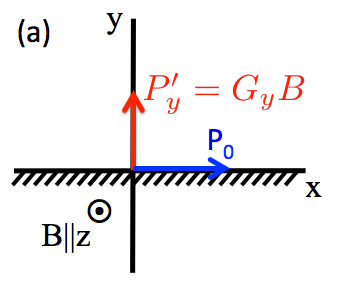}
\includegraphics[width=1.4in]{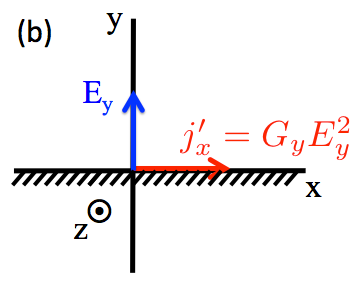}
\caption{\label{exg} (color online) Magnetoelectric Polarization (panel (a)) and electric nonlinear anomalous Hall effect (panel (b)) in a 2D system with a mirror line along x axis. In panel (a), the mirror symmetry requires the zeroth order polarization ${\bm P}_0$ to be along the mirror line, and requires the first order $\bm P^\prime$ to lie in the perpendicular direction.  In panel (b),  the linear anomalous Hall current vanishes due to the mirror symmetry, but the nonlinear anomalous Hall current can exist along the mirror line if the electric field is applied along the perpendicular direction.}
\end{figure}

Since the topological part (the first term in Eq.(\ref{me})) is quantized and well understood \cite{QHZ, Zhang2011, Kane2010, Vish2010, Bern2011, Fu2007}, we focus on the magnetoelectric polarization due to the positional shift, which requires broken time reversal and spatial inversion symmetry \cite{Vander2010}. To show its connection with the nonlinear anomalous Hall effect discussed later, we consider the two band model in Eq.(\ref{2band}), in which the second term in Eq.(\ref{me}) for the lower band gives
\begin{equation}\label{pl}
P_i^\prime=\int {d^3 k\over (2\pi)^3} G_i B\,,
\end{equation}
where $\bm G=(\hat{\bm z}\cdot\bm \partial h_0\times \bm \partial n_j)\bm \partial n_j/(4h)$, with $\hat{\bm z}$ being the direction of the magnetic field. We note that, if $h_0$ is a constant, $\bm G$ would vanish. This is consistent with previous observation that a non-zero orbital magnetoelectric polarization must require particle-hole symmetry breaking of the system \cite{Vander2010}. A minimal lattice model that realizes this effect can be constructed in 2D. Notice that for model Eq.(\ref{2band}) in 2D the topological part of magnetoelectric polarization, i.e. the first term in Eq.(\ref{me}) vanishes \cite{Xiao2009} hence only the contribution from $\bm a'$ exists. Moreover, since $\bm a^\prime$ transforms as a spatial vector and it must lie in the plane, in general it must vanish if the system has in-plane rotational symmetry. And if in-plane mirror symmetry exists, $\bm P'$ would be restricted to be along the normal direction of the mirror line (see Fig.\ref{exg} (a)). These symmetry constraints provide guidance for the construction of the lattice model, as discussed in the supplemental materials \cite{suppl}.

{\it Nonlinear anomalous Hall effect.}---
In the semiclassical approach the transport current is given by $\bm j=-\int\dot{\bm r}_c f(\bm k) \mathcal{D} \,d^3 k/(2\pi)^3\,,$ where $f(\bm k)$ is the distribution function. Because our theory is accurate up to second order in external fields, it allows us to evaluate the nonlinear current response. Here we focus on the intrinsic contribution to the Hall current which is purely from the band structure effects without disorder scattering \cite{Nagaosa2010, KL1954, Streda2010, Streda2010b, Fivaz1969, JNM2002, Onoda2002}. Under fixed temperature and uniform electromagnetic fields, we obtain the intrinsic current $\bm j^\prime$ that is second order in external fields:
\begin{align}\label{cur}
\bm j^\prime
&=\bm E\times \int\left[\bm v\times \bm a^\prime+\bm{\mathit{ \Omega}} (\bm B\cdot \bm m)\right]{\partial f_0 \over \partial \varepsilon_0}\,{d^3 k\over (2\pi)^3}\,.
\end{align}
Or more explicitly we can write
\begin{align}
\label{EE}{\partial^2 j_i^\prime \over \partial E_j \partial E_\ell}&=\int (v_{i} F_{j\ell}-v_{j}F_{i\ell}) {\partial f_0\over \partial \varepsilon_0} {d^3 k\over (2\pi)^3}\,,\\
\label{EB}{\partial^2 j_i^\prime \over \partial E_j \partial B_\ell}&= \int (v_{i} G_{j\ell} -v_{j}G_{i\ell}+\epsilon_{ijk}\mathit \Omega_{k} m_\ell) {\partial f_0\over \partial \varepsilon_0} {d^3 k\over (2\pi)^3},
\end{align}
where $f_0$ is the equilibrium Fermi-Dirac distribution function, and $\bm m=-{1\over 2} {\rm Im}\langle \bm \partial u_0|\times (\varepsilon_0-\hat{H}_c)|\bm \partial u_0\rangle$ is the orbital magnetic moment \cite{Xiao2010}. The effects associated with the two response functions in Eq.(\ref{EE}) and (\ref{EB}) shall be termed as the electric nonlinear anomalous Hall effect and the magneto nonlinear anomalous Hall effect respectively. These response functions can be directly evaluated by first principle methods.
From Eq.(\ref{cur}) we see that the intrinsic nonlinear current is purely of Hall type, and the appearance of $\partial f_0/\partial \varepsilon_0$ shows that it is a Fermi surface property. The second term in the square bracket of Eq.(\ref{cur}) comes from a correction of the band energy which could be envisioned from an naive extension of the first order theory and has been discussed in the study of anomalous Hall transport in multi-valley systems \cite{Cai2013}. The first term comes from the correction of the Berry curvature due to the positional shift found in this work, which is quite nontrivial. Moreover, we note that for in-phase oscillating $\bm E$ and $\bm B$ fields, the first order intrinsic anomalous Hall response vanishes upon time average, hence the DC intrinsic anomalous Hall current would be dominated by the nonlinear response $\bm j'$.

First let us consider the electric nonlinear anomalous Hall effect with $\bm B=0$. Then the intrinsic nonlinear Hall conductivity $\sigma_{xy}^\prime=\partial j_x/\partial E_y$ is proportional to the electric field and only the term with positional shift in Eq.(\ref{cur}) contributes. For the generic two band model (\ref{2band}), the result is
\begin{equation}
\sigma_{xy}^\prime=-\int{d^3 k\over (2\pi)^3}{\partial f_0\over \partial \varepsilon_0}\bm G\cdot \bm E\,.
\end{equation}
Interestingly, it also involves the $\bm G$ vector found for the orbital magnetoelectric polarizability. In fact, the two effects have the same symmetry properties, requiring both time reversal and spatial inversion symmetries to be broken in the system. This nonlinear anomalous Hall current will dominate if the corresponding linear current vanishes due to symmetry constraint. For example, if a 2D system has a mirror line perpendicular to the electric field, then the linear intrinsic current vanishes because the (unperturbed) Berry curvature has a sign change under mirror operation, while the nonlinear current could be finite (see Fig.\ref{exg} (b)).

In comparison, the magneto nonlinear anomalous Hall effect does not have such a stringent symmetry constraint. In fact, since the current transforms in the same way as the product of electric and magnetic fields under both time reversal and spatial inversion, this is much easier to realize in real systems. Furthermore, if the system itself has time reversal symmetry (neglect the small Zeeman splitting due to the external magnetic field), both the linear anomalous Hall effect and the electric nonlinear anomalous Hall effect vanish, and the magneto nonlinear anomalous Hall effect dominates. For the two band model, we find that
\begin{align}\label{ahe}
\sigma_{xy}^\prime&=\int {d^3 k\over (2\pi)^3} {\partial f_0\over \partial \varepsilon_0}\left[\frac{g_{ij}}{4h}(\hat{\bm z}\times \bm v)_i\,(\bm B\times \bm \partial h_0)_j \right.\notag\\
&\;\left.-{1\over 8}(\hat{\bm z}\times \bm v)_i \epsilon_{k \ell j}B_{k}\mathit \Gamma_{j \ell i}+h(\bm{\mathit{\Omega}}\cdot \hat{\bm z})(\bm{\mathit{\Omega}}\cdot \bm B)\right]\,,
\end{align}
where $\bm v=\bm \partial (h_0-h)$ and $\bm{\mathit\Omega}={1\over 2}\epsilon_{ijk}n_{i}\bm \partial n_{j}\times \bm \partial n_{k}$ is the unperturbed Berry curvature. As a concrete example, let's consider a 2D gapped Dirac model with $h_0=0$ and $\bm h=(vk_x,vk_y,\Delta)$, which is widely used to study systems such as symmetry-breaking graphene, ${\rm MoS_2}$, topological insulator surfaces with time reversal symmetry breaking, and topological insulator thin films \cite{NG2009, Xiao2012, Zhang2011, Kane2010}. Here $v$ is the fermi velocity and $\Delta$ is the gap parameter. Consider an in-plane electric field and an out-of-plane magnetic field, we obtain that (at zero temperature)
\begin{equation}\label{ahe1}
\sigma_{xy}^\prime=-e^3{v^2(v^2p_F^2+2\Delta^2)\over 16\pi(v^2p_F^2+\Delta^2)^{2}}B\,,
\end{equation}
where $p_F=\hbar k_F$ being the Fermi momentum and we assume the Fermi level is in the upper band. We have recovered factors $e$ and $\hbar$ in Eq.(\ref{ahe1}). We point out that for ${\rm MoS_2}$ or graphene (with inversion symmetry breaking) with two inequivalent valleys $K$ and $K'$ connected by time reversal symmetry, the contributions to this magneto nonlinear anomalous Hall effect from the two valleys in fact add together rather than cancel each other as in the first order response \cite{Xiao2007}.

Besides this effect, note that due to magnetic field, there is also the ordinary Hall response due to Lorentz force. Both effects have linear $B$ dependence in the transport coefficient. However there is an important difference. The ordinary Hall conductivity is proportional to the square of relaxation time (or longitudinal conductivity), while our intrinsic nonlinear conductivity does not have such extrinsic dependence. On the other hand, in terms of the Hall resistivity, the ordinary effect has an intrinsic looking form $\rho_{xy}^\text{ord}=-B/ne$, where $n$ is the carrier density, while intrinsic nonlinear Hall resistivity acquires a dependence on the square of the longitudinal resistivity. This is well understood as a result of matrix inversion between the conductivity and resistivity tensors with the usual condition that longitudinal coefficients are much bigger than the transverse ones. Therefore from this discussion, like the linear anomalous Hall effect, our nonlinear effect would become important for more resistive samples.

The argument above can be quantified for the gapped Dirac model by calculating the ratio between the two contributions explicitly. From Eq.(\ref{ahe1}), we have 
\begin{equation}
{\rho_{xy}^\prime\over \rho_{xy}^\text{ord}}=\left(\rho_{xx}{e^2\over 4h}\right)^2\left[1-\left({\Delta\over \varepsilon_F}\right)^4\right]\,,
\end{equation}
where $\varepsilon_F=\sqrt{v^2p_F^2+\Delta^2}$ is the Fermi energy, $\rho_{xy}^\prime\simeq \sigma'_{xy}\rho_{xx}^2$ is obtained by inverting $\sigma_{xy}'$, and $\rho_{xx}$ is the longitudinal resistivity. Not surprisingly, the first factor is proportional to $\rho_{xx}^2$. The second factor shows a simple dependence on Fermi energy: it vanishes at the band bottom and quickly saturates with increasing Fermi energies. Our predictions can be tested by the standard Hall bar measurement on single layer ${\rm MoS_2}$ or ${\rm Bi_2Se_3}$ thin films with carrier density tuned by doping or electric gating techniques.

{\it Summary.}---We have derived the gauge-invariant positional shift of wave packet center due to external fields, based on which a complete second order semiclassical theory has been constructed. The positional shift modifies the non-canonical geometry of Bloch electrons. As applications, we show that the full expression of the orbital magnetoelectric polarizability can be easily derived by our theory. We further predict that the positional shift leads to a nonlinear Hall response which can dominate under certain symmetry constraints.

{\it Acknowledgment.}---We acknowledge useful discussions with Z. Qiao, H. Chen,  J. Zhou, X. Li, R. Cheng, and L. Zhang. YG is supported by the MOST Project of China (2012CB921300), and NSFC (91121004). SAY is supported by SUTD-SRG-EPD-2013062. QN is supported by DOE (DE-FG03-02ER45958, Division of Materials Science and Engineering) and Welch Foundation (F-1255).


\begin{thebibliography}{99}
\bibitem[8]{foot1} Blount systematically classified the contribution to the effective Hamiltonian Eq.(I.(30)) in Reference [5]. However, the effective Hamiltonian is gauge dependent since the third term $\bm a^s\cdot \bm v_A$ depends on the gauge choice while all the other terms are gauge invariant.
\end{thebibliography}
\end{document}